# Plasmonic graphene photodetector based on channel photo-thermoelectric effect

*Jacek Gosciniak[1], Mahmoud Rasras[1], and Jacob Khurgin[2]*

[1]*New York University Abu Dhabi, Saadiyat Island, P.O. Box 129188, Abu Dhabi, UAE.*
[2]*John Hopkins University, Baltimore, MD 21218, USA*

**Abstract**
We propose an on-chip CMOS compatible graphene plasmonic photodetector based on the photo-thermoelectric effect (PTE) that occurs across an entire homogeneous graphene channel. The proposed photodetector incorporates the long-range dielectric-loaded surface plasmon polariton (LR-DLSPP) waveguide with a metal stripe serving simultaneously as a plasmon supporting metallic material and one of the metal electrodes. Large in-plane component of the transverse magnetic (TM) plasmonic mode can couple efficiently to the graphene causing large temperature increases across an entire graphene channel with a maximum located at the metal stripe edge. As a result, the electronic temperatures exceeding 12000K at input power of only a few tens of µW can be obtained at the telecom wavelength of 1550nm. Even with limitations such as the melting temperature of graphene (*T*= 4510 K), a responsivity exceeding at least 200 A/W is achievable at telecom wavelength of 1550 nm. It is also shown that under certain operation conditions, the PTE channel photocurrent can be isolated from photovoltaic and p-n junction PTE contributions providing an efficient way for optimizing the overall photodetector performance.

**Introduction**
Photodetectors (PDs) are the essential building blocks of an optoelectronic links that convert light into an electrical signal. The monolithic, on-chip, optoelectronic integration requires development of CMOS-compatible PDs operating in the telecom wavelengths (1.1 - 1.7 µm) based on CMOS technology [1-4]. Although sensitivity is the most important attribute for photodetectors in long distance communications, for short distance interconnects the most critical factor is the total energy dissipated per bit. The optical energy received at a photodetector is directly related to transmitter optical output power and the total link loss power budget, which includes total link attenuation, coupling losses, and eventually, a power margin. Hence, for 10 fJ/bit transmitted optical energies, the received optical energy would be 1 fJ/bit [3, 4]. Thus, minimizing the optical losses at the photodetector is crucial for overall performance of the system.

Traditionally, in silicon photonics, the photodetectors are based on silicon (Si) [5], germanium (Ge) [6, 7] or III-V compound semiconductors [8, 9]. All of them, however, have a number of shortcomings. In terms of Si photodetectors, the photon energies at telecom wavelength are not sufficient for direct photodetection [5]. To overcome this limit, Schottky photodetectors based on doped Si were proposed with responsivity exceeding 0.5 A/W at the telecom wavelength of 1550 nm [10], however these detectors have large dark current. While III-V detectors have excellent operational characteristics, the direct monolithic integration of III-V photodetectors on Si wafers remains a challenge because of the large lattice constant mismatch and different thermal coefficients [8, 9]. Similar problems are associated with Ge photodetectors [6, 7]. For this reason, the search for better materials is essential for the progress in high-speed photodetectors.

Graphene is a very attractive material for photonic and optoelectronics because if offers a wide range of advantages compared to other materials [11, 12]. Single-layer graphene absorbs 2.3 % of incident light, which is remarkably high for an atomically thin material. Graphene is gapless which enables charge carrier generation by light absorption over a very wide energy spectrum. It has ultrafast carrier dynamics, wavelength-independent absorption, tunable optical properties, high mobility, and when used in infrared the ability to confine electromagnetic energy to very small volumes [12, 13]. The high carrier mobility (both electrons and holes) enables ultrafast conversion of photons or plasmons to electrical currents or voltages [14]. By integration with local gates, this process is tunable in situ and allows for submicron detection resolution [15, 16]. Moreover, graphene offers low-cost manufacturing based on complementary metal-oxide-semiconductor (CMOS) fabrication processes allowing integration on wafer-scales [17, 18]. By integration of graphene with the propagating photonic mode, the interaction length of the mode with graphene can be greatly enhanced compared to the situation with normal incidence in which the interaction length is limited by the thickness of graphene [19-22]. The graphene photodetectors have the additional advantage of process compatibility with standard photonic integrated circuits [22-24]. However, up to date, most of the graphene photodetectors suffer from either low responsivity [14, 17, 19, 21, 22, 40], large dark currents associated with biasing the graphene channel [20, 21, 39] or complex design [26, 39]. Thus, finding a better photodetector configuration that can takes full advantages of graphene's properties is highly desired.

1. **Photo-thermoelectric effect**

To find an answer to the questions raised above, we first need to classify the main effects that can contribute to photocurrent (or photovoltage) generation in waveguide-integrated graphene devices [25-28]. One can distinguish three main effects – photo-voltaic (PV) [29-34], photo-bolometric (PB) [35-39] and photo-thermoelectric (PTE) [40-43] effects. The choice of the effect depends on device configuration, design and operation conditions [25, 28, 34, 40]. Recent studies have shown that the PTE effect dominates over PV photocurrent generation in graphene devices [44, 45], therefore it is the PTE devices that we consider optimizing in this work.

The operational principle of most PTE detectors relies on a temperature gradient in the material's p-n junction under an electromagnetic field excitation, resulting in a net thermoelectric voltage across the material due to the Seebeck effect [16, 21, 26, 33, 40-43, 46]. Thus, the generated photocurrent is proportional to the Seebeck coefficient for the two sides of the junction. As a result, materials with a high Seebeck coefficient and/or able to sustain a high temperature are highly desired. However, most thermoelectric materials suffer from either low Seebeck coefficients or low melting temperatures that prevents them from reaching high operation temperature, consequently limiting the the PTE photodetector's performance [47]. Additionally, they suffer from the relatively long response time originating from the phonon-dominated transport, which is typically on the order of milliseconds [47].

Graphene is a very attractive material for such a photodetector. Apart from the excellent properties as outlined in the introduction, it can sustain a very high temperature [48], thus, enhancing a photocurrent [47]. Furthermore, graphene PTE detector relies upon the gradient of electron temperature $T_e$ rather than lattice temperature and electron temperature can be higher than lattice temperature. The response time is dominated by the electron-phonons interaction and is in the range of 2-4 picoseconds [49-52]. Operation of such a photodetector requires a spatially inhomogeneous doping profile that is created by local heating of p-n junction by incident laser power [16, 21, 40-43]. As a result, the nonequilibrium hot carriers are excited with an electron temperature $T_e$ higher than that of the lattice, giving rise to the electron temperature gradient across a junction. This increase in the temperature of photoexcited carriers is a direct consequence of the large optical phonon energy in graphene (~200 eV) and the low scattering rate of electrons by acoustic photons [50-54]. The latter give rise to an increased temperature of the photoexcited carriers for picoseconds, while the lattice stays close to room temperature. In this photodetector, the photovoltage is generated from hot electrons through the Seebeck coefficient that varies in the graphene sheet as a result of different doping or temperature gradient. For different doping levels on both sides of the junction and the temperature profile in the graphene, the Seebeck coefficient varies across the graphene generating a photo-thermoelectric voltage. Due to the non-monotonic nature of the difference in Seebeck coefficients in the two differently doped regions of the junction, the resulting photovoltage exhibits multiple sign reversals in dependence of the gate voltage. The characteristic 6-fold pattern is the result of the two doping levels on either side of the junction [14, 15, 25, 26, 40-42]. Another approach relies on heating of one contact that can lead to a temperature gradient across a graphene channel resulting in a photo-thermoelectric contribution to the photovoltage [40, 55].

However, most of these graphene photodetectors suffer from low carrier's temperature which make their contribution to the overall photocurrent negligible [19, 21, 24, 51]. To enhance a photocurrent, either materials with higher Seebeck coefficient or higher temperature gradient in the material are required. Due to the superior transport properties of graphene, many efforts now focus on the efficient conversion of power to the electronic temperature of graphene and, simultaneously, on a high temperature gradient in graphene.

## 2. Proposed graphene photodetector arrangement

First to ensure a high power conversion efficiency on the electronic temperature in graphene, it is essential to efficiently couple a light to a photodetector. This can be achieved in a waveguide-integrated arrangement that provides a high coupling efficiency, *i.e.*, most of the power coupled to the photodetector can participate in photocurrent generation. Furthermore, most of the power in a photodetector should be absorbed by the graphene while absorption losses in metal should be avoided. To achieve this, a strong in-plane electric field component of the mode is required. Additionally, a high electron temperature gradient in graphene should be achieved. In practice this means that the in-plane component of the electric field in graphene should be strong but also decay very fast from one electrode to another. Taking into account those aspects, we proposed here a photodetector that is based on the metal-graphene-metal (MGM) arrangement [29, 30, 43] with a graphene channel contacted by two electrodes, either of the same [30] or two different metals [29, 34]. The difference in work function between the metal pads and graphene leads to charge transfer with a consequent shift of the graphene's Fermi level that is in contact with the metal [29, 34, 56]. As the Fermi level of the metal-free graphene channel is different, this results in a chemical potential gradient between the metal-free graphene and graphene in contact with metal [29, 34] and accordingly a gradient of carrier concentration. This inhomogeneous carrier concentration (or effective doping) profile [55, 56] creates a graded junction along the channel that is very important in the photodetection process, as it results in an internal electric field capable of separating the light induced e-h pairs.

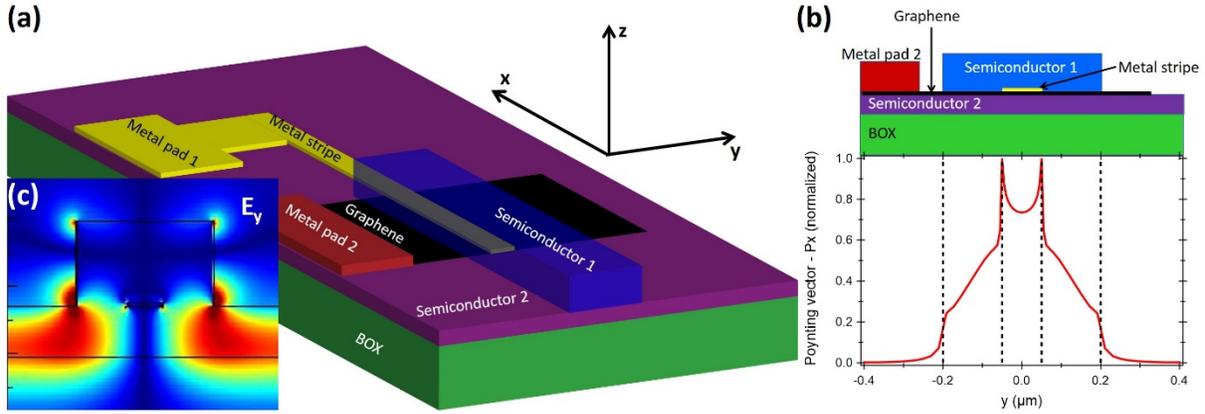

**Fig. 1.** (a) Schematic of the proposed graphene photodetector with a photonic waveguide integrated with the photodetector on the same material platform. Here Semiconductor 1 and Semiconductor 2 are taken as Si. (b) Cross-section through the structure with the Poynting vector $P_X$ showing a field distribution in a direction of the external electrode (Metal pad 2). (c) In-plane electric field component of the LR-DLSPP mode.

Proposed waveguide-integrated graphene photodetector can provide, a high electron temperature in graphene under low excitation power, an extremely high temperature gradient in graphene, a short graphene channel length and flexibility in design. Figure 1a shows schematic of the proposed graphene photodetector realized in a long-range dielectric-loaded surface plasmon polariton (LR-DLSPP) waveguide configuration [58-61] with a plasmonic mode guided by a metal stripe serving simultaneously as one of the electrodes (Fig. 1b). The electric field of the propagating mode is highly localized between metal and semiconductor leading to enhanced absorption in graphene sheet placed below the metal stripe (Fig. 1b). Importantly, the in-plane electric field component of the LR-DLSPP modeessential for absorption in graphene, is very strong close to the graphene sheet even for the TM mode (Fig. 1c) [58-62]. In our design, only half of the waveguide shown in Fig. 1b is needed for photodetection purpose as presented in Fig. 2. As shown in Fig. 2, the maximum temperature is located at the edge of the metal stripe ("hot" electrode) while the external electrode is not affected by the electric field and its temperature is at a minimum ("cold" electrode). As a result, temperature decays in the direction of the external electrode with the cooling rate and cooling length of hot carriers in graphene depending mostly on the width of neutrality region $\Delta$ and chemical potential $\mu$.

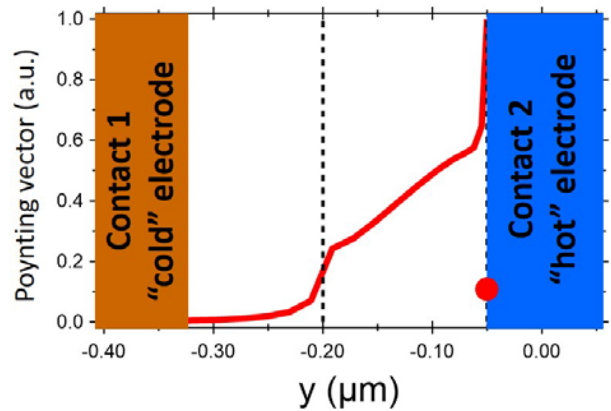

**Fig. 2.** A proposed PTE channel photodetector with the Poynting vector $P_X$ with the field distribution between metal stripe (Contact 2 -"hot" electrode) and the external electrode (Contact 1 - "cold" electrode).

Since the electron-electron scattering in graphene is ultrafast, on the order of 10 fs, and electron-phonon scattering relatively slow, on the order of picoseconds [49-54], the photoinduced carriers are thermalized with an effective temperature $T_e$ before they dissipate heat to the lattice. Typically the cooling length of hot electrons in graphene exceeds hundreds of nanometers, and so, the electrons in graphene would not reach thermal equilibrium with the lattice before being collected [26, 49-52].

As mentioned earlier, in the proposed design the electric field reaches its maximum at the metal stripe edge ("hot" electrode) and quickly decays in the direction of second electrodemleading to an asymmetric temperature distribution that can be transduced into an electrical signal through the Seebeck effect. Even more essential to the excitation of currents in graphene, the in-plane electric field component ($E_y$) of the LR-DLSPP mode is very strong close to the graphene surface, even for the TM supported mode [58-61], due to the sharp metal stripe corners (Fig. 1c) [62]. The decay length of the $E_y$ electric field on the graphene in the direction of the external electrode is extremely small in the range of a few nanometers (Fig. 1c). This produces an extreme temperature gradient in the graphene which enhances the photodetector performances.

### 3. Estimate of power absorbed by graphene sheet

To evaluate the performance of the proposed photodetector the amount of power absorbed by a graphene sheet need to be determined.

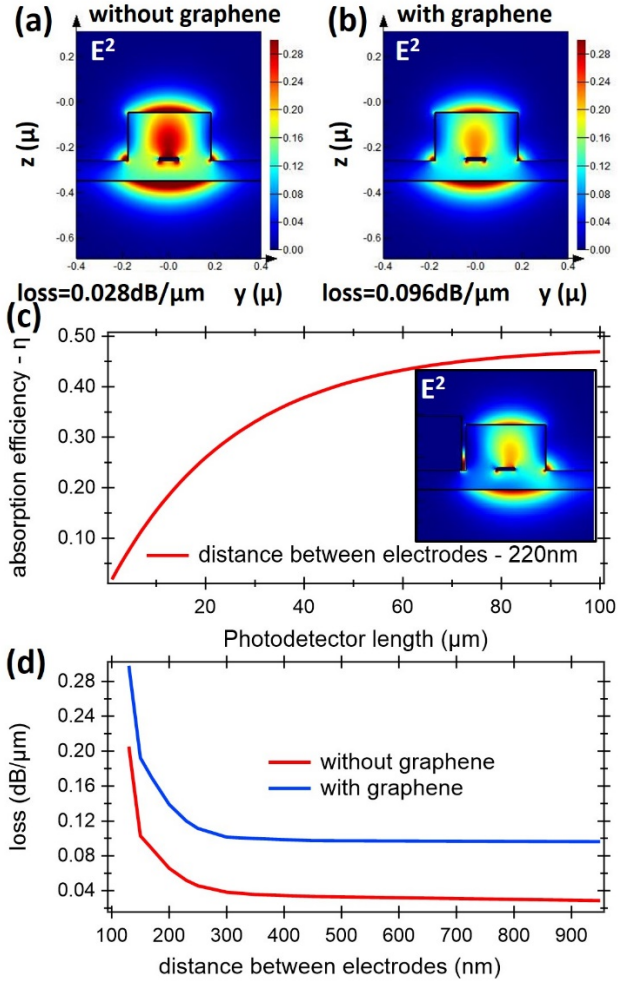

**Fig. 3.** (a), (b) Calculated $E^2$ of the LR-DLSPP waveguide mode with corresponding losses for structure without graphene (a) and with graphene (b). (c) Calculated absorption efficiency versus photodetector length for a photodetector with 220 nm separation distance between electrodes. (d) Propagation losses versus distance between electrodes for structures without graphene and with graphene.

Power absorbed by the graphene sheet can be calculated as follow:

$$P_{abs} = P_{in}e^{-\alpha \cdot L} \quad (1)$$

Here we simulate around $P_{in}$=165 µW of power coupling to a photodetector. For the absorption coefficient of graphene $\alpha_G$ and metal $\alpha_M$ obtained from a simulation (Fig. 3a, b), the length-dependent of fraction of light absorption in graphene $\eta$ can be calculated by:

$$\eta = \frac{\alpha_G}{\alpha_M + \alpha_G}(1 - e^{-\alpha_G L}e^{-\alpha_M L}) \quad (2)$$

where $L$ is the length of photodetector. Calculations were performed for a telecom wavelength of 1550 nm and results were summarized in Fig. 3c. From this figure, it can be deduced that for a 40 µm-long photodetector about 40 % of the power is absorbed by graphene (Fig. 3c). However, only half of this power will contribute to the photodetector's performance. As a result, we can assume a reasonable power of $P_{abs}$=33 µW absorbed by graphene that will contribute to a photocurrent generation.

The limit of the absorption coefficient of graphene $α_G$ in a given waveguide configuration can be calculated from a simple formula that relate a 2D graphene absorption $α_{2D}$ with the effective thickness of the waveguide $t$ through the expression:

$$α_G = α_{2D}/t \qquad (3)$$

Taking $α_{2D}$=2.3 % [11-13] and the effective thickness of the waveguide at $t$=100 nm, the absorption of graphene was calculated at $α_G$=0.23 dB/µm. Thus, to achieve 50 % power absorption by graphene, a 13 µm-long waveguide/photodetector is required. As shown, further improvements of graphene-based photodetector performances are possible.

### 4. Channel PTE current generation principle

Most previously reported PTE effects usually refer to the junction formed either by monolayer and bilayer graphene or between regions of graphene with different Fermi energies $E_F$, e.g. p-n junctions with buried split-gates or with a top-gated control. As previously demonstrated, the main utility of the PTE effect is to creating an asymmetry in the device. Such an asymmetry can be created by using two different contact metals [29] or by using two adjacent graphene regions of different doping [26].

The photovoltage in the present case is generated at the junction and is driven by the difference in graphene's Seebeck coefficient $ΔS=S_1-S_2$ on either side of the junction through:

$$V_{ph} = ΔSΔT = (S_1(μ_1) - S_2(μ_2))ΔT \qquad (4)$$

where $ΔT$ is the electron temperature increase within the junction after photoexcitation and $μ$ is the chemical potential. The photocurrent is then determined as:

$$I_{ph} = \frac{ΔS}{R}ΔT \qquad (5)$$

which results in multiple photocurrent sign reversals over a gate voltage sweep due to the non-monotonic dependence of $S_1$ and $S_2$ on µ.

Unlike PTE effect, the photovoltaic (PV) effect relies on the separation and then collection of photoinduced electrons and holes by a built-in electric field leading to a net photocurrent:

$$I_{pv} = \frac{μ_mΔ}{σ_0R}\left(\tan^{-1}\frac{μ_1}{Δ} - \tan^{-1}\frac{μ_2}{Δ}\right)n_z^{ave}(y=0) \qquad (6)$$

where $μ_m$ is the carrier's mobility, $Δ$ the width of the neutrality region of the graphene channel, $σ_0$ the minimum conductivity, $R$ the total resistance, and $n_x$ the steady state density of photoexcited carrier. Here, $μ_1$ and $μ_2$ suggest a chemical potential shift that results from different doping levels of two opposite contacts introduced by different metals (see section 6). Under light excitation, the hot carriers are formed on a timescale on the order of 100 fs, followed by a slow picosecond carrier recombination and cooling. The built-in electrical field in the channel separate photogenerated carriers and facilitate their drift to the contacts. In consequence, a net photocurrent is generated without the application of a bias voltage.

From Eq. 6 it can be deduced that photoresponse reaches maximum in the vicinity of a p-n junction when $μ_1$ and $μ_2$ are opposite sign. As the polarity of the PV current is determined solely by the sign of field gradient, there is only one sign reversal occurring at $μ_1=μ_2$. Thus, the photovoltage induced by the PV effect changes monotonically with the gate voltage.

Here we focus, however, on the graphene-metal (G-M) interface where the chemical potential of graphene that is in contact with metal is shifted compared to the graphene channel that is related to the difference in work functions of the materials [55-57]. In the proposed arrangement, the temperature difference is established across the entire device channel. It provides a few advantages over a narrow p-n junction, as more of the electron heat can be converted into a photovoltage giving rise to a responsivity increase [20, 21].

In contrast to the p-n junction PTE, the channel PTE voltage is driven by temperature across the entire channel as:

$$V_{ph} = S\Delta T = S(\mu)\Delta T \tag{7}$$

The Seebeck coefficient *S* of the graphene channel is defined as:

$$S(\mu) = -\frac{\pi^2 k_B^2 T}{3e}\frac{1}{\sigma}\frac{d\sigma}{d\mu} \tag{8}$$

where $k_B$ is the Boltzmann constant, *T* is the lattice temperature, *e* is the electron charge, $\sigma$ is the conductivity of graphene, and $\mu$ is the chemical potential. Consequently, the channel PTE exhibits a single sign change in the channel due to the monotonic dependence of *S* on $\mu$.

The operating principle of the proposed LR-DLSPP graphene photodetector is as below:

The propagating plasmonic mode reaches maximum in the metal-semiconductor interface (Contact 2) and decays fast in the direction of an external electrode (Contact 1) (Fig. 1b and Fig. 2). The strongly enhanced electromagnetic field from the plasmonic waveguide highly improves photo-absorption in the nearby graphene, resulting in efficient and localized carrier heating in the graphene. These carriers thermalize rapidly leading to local electron temperature increase. The photoexcited hot electrons diffuse in the direction of the external electrode (Contact 1) and create a potential gradient *ΔV=-S(y)∇T$_e$(y)*, where *∇T$_e$(y)* is the temperature gradient of the hot electrons and *S(y)* is the Seebeck coefficient. As a results, the photocurrent is collected between the source (Contact 1) and drain (Contact 2) that is given by:

$$I_{ph} = \frac{S}{R}\Delta T \tag{9}$$

where *R* is the resistance of the graphene sheet:

$$R = \frac{L}{2W}\left(\frac{\sigma_1+\sigma_2}{\sigma_1\cdot\sigma_2}\right) \tag{10}$$

Assuming *L*=400 nm spacing between electrodes, photodetector length *W*=40 μm, and a photodetector operating based on the G-M channel photo-thermoelectric effect ($\sigma_1=\sigma_2$), the graphene sheet resistance was calculated at *R*~50 Ω at chemical potential *μ*=0 eV.

5. **Temperature distribution**

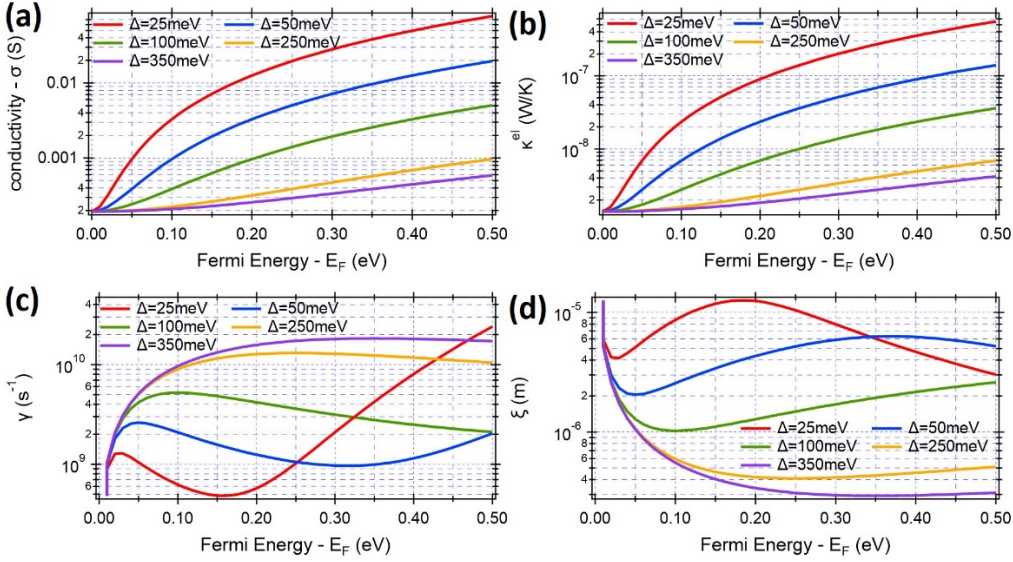

**Fig. 4.** Calculated physical parameters for graphene with different charge neutrality points *Δ*. Calculated conductivity *σ* (a), electron thermal conductivity *κ*$^{el}$ (b), cooling rate *γ* (c), and cooling length *ξ* (d) of graphene with *σ*$_0$ = 5(e²/h), and *Δ*=25, 50, 100, 250 and 350 meV, respectively. Calculations were performed with electron-phonon coupling strength *g*≈0.2294.

The electron temperature of hot electrons in graphene is governed by the heat transfer equation [26, 49, 50]:

$$\kappa^{el}\nabla^2 T_e - \gamma C^{el}(T_e - T_0) + P^* = 0 \qquad (11)$$

where *κ*$^{el}$ represents the lateral 2D thermal conductivity and *γC*$^{el}$ the vertical heat loss, $T_e$ is the electron temperature at a given position, $T_0$ is the temperature of the substrate and P* is the input power density provided by the LR-DSLPP mode. Thermal conductivity *κ*$^{el}$ can be calculated from the Wiedemann-Franz relation [50] and expressed by:

$$\kappa^{el} = \frac{\pi^2 k_B^2 T}{3e^2}\sigma \qquad (12)$$

where $k_B$ is the Boltzman constant, *e* is the electron charge, *T* is the operating temperature and *σ* is the electrical conductivity of graphene. The 2D electrical conductivity of graphene depends on the chemical potential/ Fermi level shift and is expressed by:

$$\sigma(\mu) = \sigma_0\left(1 + \frac{E_F^2}{\Delta^2}\right), \quad \sigma_0 = 5\left(\frac{e^2}{h}\right) \qquad (13)$$

where *σ*$_0$ denotes the minimum conductivity and *Δ* the spreading of the transfer characteristics. Here σ$_0$ was assumed as be *σ*$_0$=0.193 mS, close to the experimentally achieved value of ~0.21 mS in Ref. 26.

As the electrons in graphene are very well thermally isolated from the lattice, the electron thermal conductivity *κ*$^{el}$ and the thermal coupling between electrons and the lattice *γC*$^{el}$ govern the heat dissipation and hence determine the electron temperature distribution. Due to the higher heat capacity of the photon system in comparison to the electronic one, the phonon system can be treated as an ideal thermal bath with $T_0$ staying constant. Here *γ* represents the electron-lattice cooling rate and *C*$^{el}$ the electron heat capacity. In graphene with a linear electronic dispersion, both parameters are very small meaning that the vertical heat dissipation is mainly limited by the electron-lattice cooling.

The distance between the position of the peak electron temperature $\Delta T_e$ and the metal contact can be characterized by the cooling length of hot electrons given by $\xi=(\kappa^{el}/\gamma C^{el})^{1/2}$. Figure 4 plots the calculated conductivity $\sigma$, electron thermal conductivity $\kappa^{el}$, electron cooling rate $\gamma$, and the electron-lattice cooling length $\xi$ of graphene for various values of the width of the neutrality region $\Delta$. Calculations were performed with electron-phonon coupling strength $g=0.2294$. As the carrier-carrier scattering in graphene is much faster than electron-phonon cooling, it makes an graphene ideal material for realizing PTE photodetectors. In such a photodetector, the whole temperature dependence of photocurrent is through $\xi$ and thus via the cooling rate $\gamma$ only. For the charge neutrality points of a graphene channel between $\Delta=250$ meV and $\Delta=350$ meV the cooling rate was calculated at 10-20 ns$^{-1}$ (Fig. 4c). Consequently, $\xi$ drops as the charge neutrality points $\Delta$ increases and reaches $\xi\approx300$ nm for $\Delta=350$ meV and Fermi energy between 300-400 eV. For a cooling length shorter than the photodetector length, the hot carriers strongly thermalize with the lattice before reaching the external electrode. However, for $\Delta<250$ meV the cooling length $\xi$ exceeds the photodetector length of 400 nm. Thus, the energy loss from the electronic system to the lattice is minimized, leading to a high temperature to drive Seebeck coefficient and giving rise to a strong PTE effect.

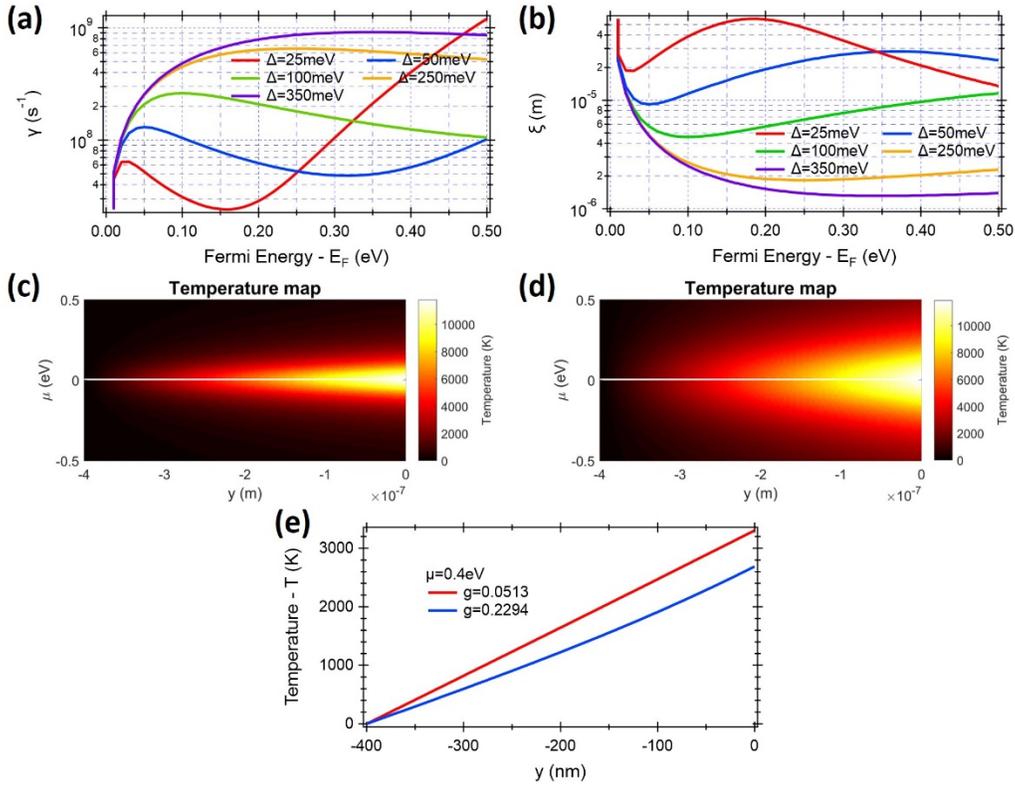

**Fig. 5.** Calculated cooling rate $\gamma$ (a), and cooling length $\xi$ (b) of graphene with $\sigma_0 = 5(e^2/h)$, and $\Delta=25, 50, 100, 250$ and $350$ meV, respectively. Calculations were performed with electron-phonon coupling strength $g\approx0.0513$. (c, d) Temperature map in a graphene for different $\Delta$: (c) $\Delta=100$ meV and (d) $\Delta=250$ meV. (e) Comparison of temperature distributions along the metal contacts for chemical potential of 0.4 eV and for different electron-phonon coupling strengths: $g=0.0513$ and $g=0.2294$.

In comparison, for the lower electron-phonon coupling strength $g=0.0513$, cooling rate exceeds 0.6 ns$^{-1}$ even for charge neutrality point $\Delta=250$ meV in the entire Fermi energy ranges from 0 eV to 0.5 eV (Fig. 5a). This results in a cooling length of $\xi=1.8$ μm, which is much longer than the photodetector length $L=400$ nm even for a reasonably high charge neutrality point Δ=250 meV (Fig. 5b). It is even higher for lower Δ exceeding 4.6 μm for Δ=100 meV.

The temperature distribution in the photodetector was calculated using the analytical solution to the heat equation:

$$\Delta T = T_e(y) - T_0 = \frac{\xi \sinh((L-|y|)/\xi)}{2 \cosh(L/\xi)} \left(\frac{P^*}{\kappa^{el}}\right) \tag{14}$$

where $P^*$ is the rate at which heat enters the system and $L$ is the photodetector length, *i.e.*, the distance between electrodes. As seen from Eq. 14, material parameters that affect the temperature profile include the thermal conductivity $\kappa^{el}$, the electronic specific heat $C^{el}$, and the electron-lattice cooling rate $\gamma$. The combination of these three parameters generates a characteristic cooling length $\xi$ for hot-carrier propagation in the system. As can be seen in Fig. 5c, d and Fig. 6b-d, temperatures reach a maximum exceeding 12,000 K at $\mu$=0 eV and decays as the chemical potential increases. The decay rate depends on Δ – for smaller Δ, the faster the decay. Simultaneously, the maximum temperature is localized at the metal stripe (Contact 2) and decays fast in the direction of the external electrode (Contact 1) (Fig. 2). When compared with the previous calculations with g=0.2294 where a maximum temperature of 2600 K was calculated for a chemical potential $\mu$=0.4 eV, it can be observed that lower electron-photon coupling strength g=0.0513 leads to higher temperature T=3200 K with the same chemical potential $\mu$=0.4 eV. In both cases the calculations were performed for Δ=250 meV (Fig. 5e).

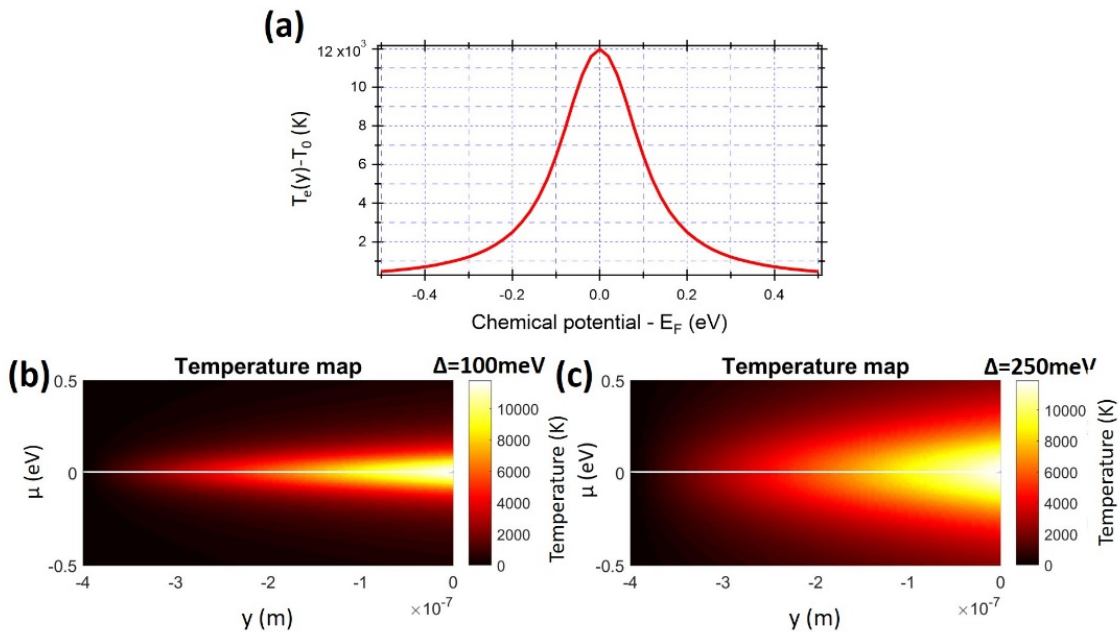

**Fig.6.** (a) Temperature rise in a graphene as a function of the chemical potential and (b), (c) temperature map in a graphene for different Δ: (b) Δ=100 meV and (c) Δ=250 meV.

6. **Photocurrent map – operation regime**

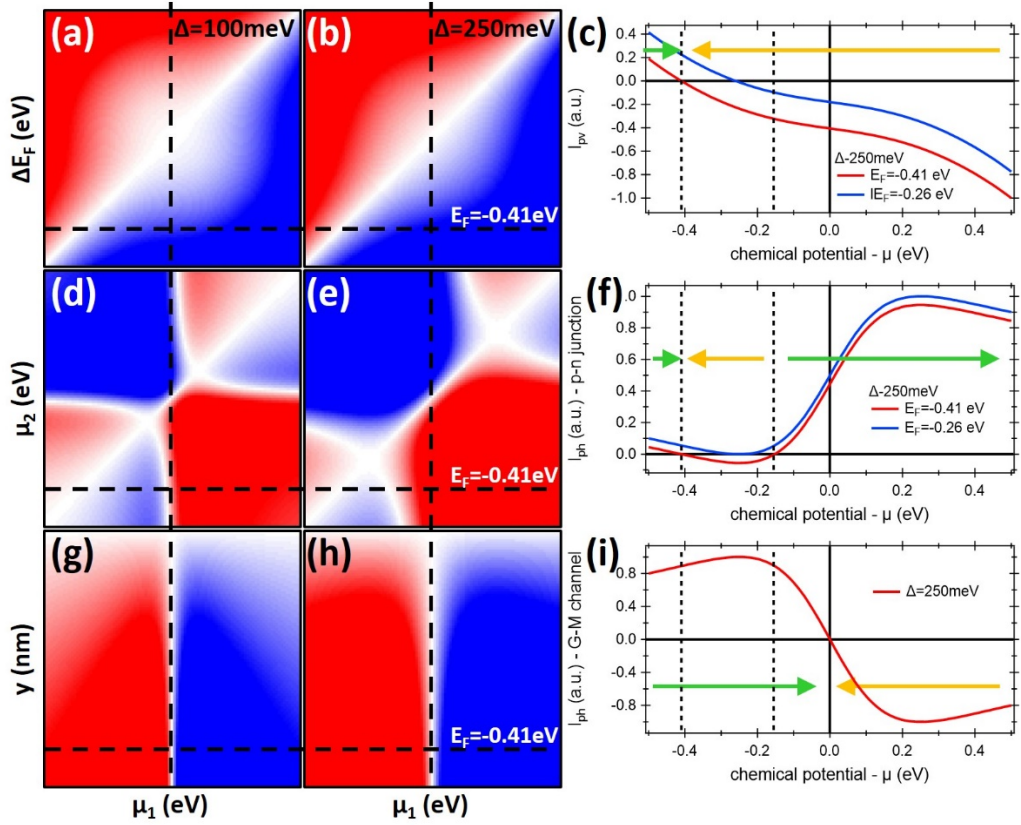

**Fig. 7.** (a)-(b), (d)-(e) and (g)-(h) Photocurrent maps of the devices operating on the basis of (a)-(b) photovoltage (d)-(e) photo-thermoelectric based on p-n junction and (g)-(h) photo-thermoelectric based on graphene-metal (G-M) channel effects. The width of the neutrality region $\Delta$ was taken at 100 meV and 250 meV. (c), (f), (i) Corresponding photocurrent versus chemical potential for (c) photovoltage, (f) photo-thermoelectric based on p-n junction and (i) photo-thermoelectric G-M channel effect for $\Delta$=250 meV. Photocurrents taken at $E_F=\mu_2$=260 eV and 410 meV (dashed line on the photocurrent maps).

The proposed photodetector can operate without the need for a bias voltage as the electronic gradient across the graphene channel is incorporated through an asymmetric electrical contact arrangement. Here one of the photodetector electrodes simultaneously serves as a metal stripe supporting a propagating mode. Thus, a highly enhanced electric field is present around this electrode with the maximum localized at the metal and decaying fast into graphene channel, in the direction of the external electrode. As a result, an electronic temperature difference is established across the channel with built-in potential difference in the channel. In Fig. 7g-h the length y and the chemical potential $\mu_1$ dependent photocurrent map is shown for different charge neutrality regions $\Delta$=100 meV and $\Delta$=250 meV. From Fig. 7g-i is clear that a photogenerated current $I_{ph}$ exhibits a single sign change at the Diract point of graphene what similar to what has been reported earlier [40, 52]. In comparison, with the p-n junction PTE the photocurrent is driven by the difference in graphene's Seebeck coefficients on either side of the junction resulting in multiple photocurrent sign reversals over a gate voltage (Fig. 7d-f). This behavior is a clear evidence of the non-monotonic dependence of $S_1$ and $S_2$ on $E_F$. The next mechanism responsible for photovoltage generation is based on the PV effect in which the photocurrent is related to $\Delta E_F$ (Fig. 7a-c). For this effect, a single sign change is observed at the flat-band point.

As observed in Fig. 7, for a chemical potential of -0.41 eV both effects, PV and junction PTE, are canceled. This potential corresponds to the flat-band condition [40], for which both the channel graphene and graphene in contact with metal are doped to the same level. The only effect contributing to a photocurrent here is the channel PTE (Fig. 7).

## 7. **Evaluating PTE performances**

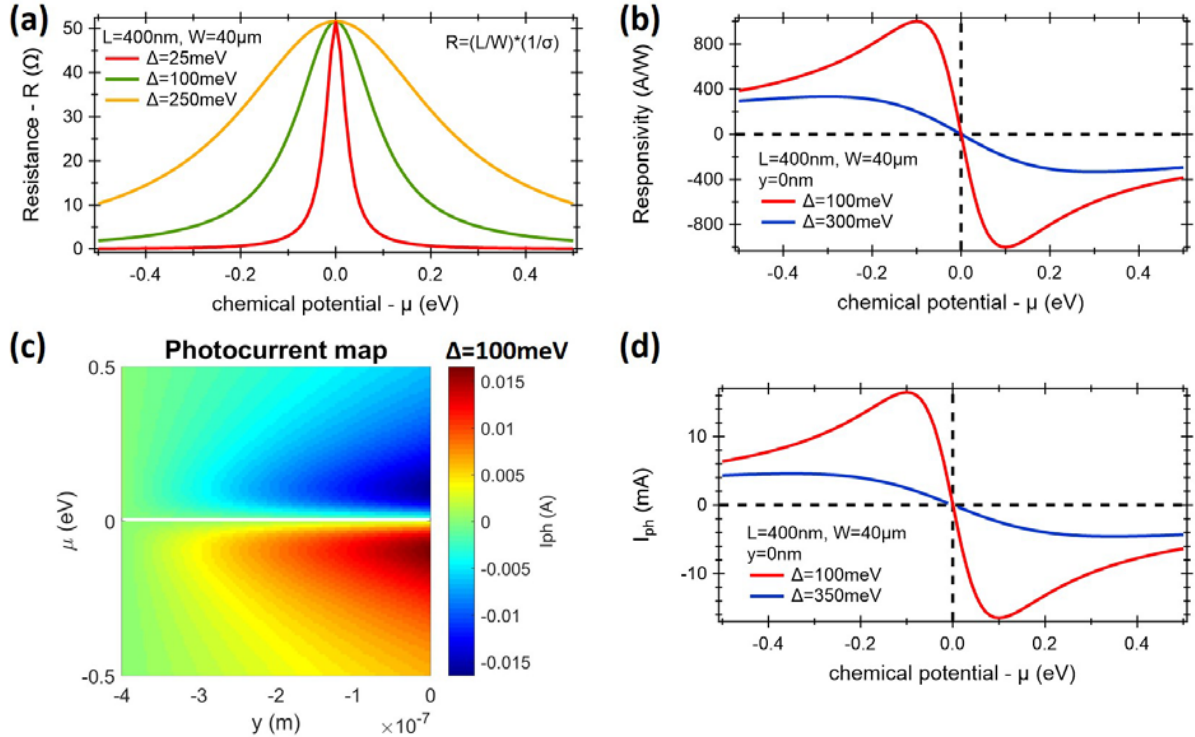

**Fig. 8.** Calculated (a) resistance $R$ of the graphene sheet, (b) photodetector responsivity $R_{resp}$ and (c), (d) photocurrent of the photodetector as a function of the chemical potential $\mu$. The length of the photodetector was kept at $L$=400 nm and width at $W$=40 µm. Calculations were performed for different charge neutrality points.

For a $W$=40 µm long photodetector and a distance $L$=400 nm between electrodes the resistance of the graphene was calculated as a function of chemical potential for different widths of the neutrality region $\Delta$ (Fig. 8a). As expected, the plot is symmetric with a maximum resistance of ~50 Ω calculated at chemical potential $\mu$=0 eV. The resistance shows a gate-dependent characteristics where doping of graphene increases its conductivity which reduces the graphene resistance. Fig. 8b shows a responsivity of the G-M photodetector at $P_{in}$=165 µW ($P_{abs}$=33 µW) for different widths of the neutrality regions $\Delta$ calculated at $R_{resp}$=1000 A/W for $\Delta$=100 meV and $R_{resp}$=350 A/W for $\Delta$=300 meV. The two sign changes of $I_{ph}$ along $\mu$=0 eV reflects the two sign change of the $S$ gradient across the junction (Fig. 8c, d). Furthermore, $I_{ph}$ is maximum close to the $\Delta$ where $S$ is largest and drop-off at higher $\mu$, *i.e.*, higher doping (Fig. 8c, d). As a result, the photocurrent calculated close to the $\Delta$ exceeds $I_{ph}$=16 mA for low level $\Delta$=100 meV and drops to $I_{ph}$=4.2 mA for larger $\Delta$=350 meV.

Even assuming a melting temperature of graphene at $T$=4510 K [63] or above $T$=6000 K as recently predicted [64], the responsivity between $R_{resp}$=100-300 A/W can be achieved at much lower power coupled to the photodetector. Thus, the photodetector can ensures the high temperature of hot carriers and associated enhanced photoresponse, while the large photodetector's width minimize the graphene sheet resistance and contact resistance.

## 8. Asymmetric metal arrangement and operation speed

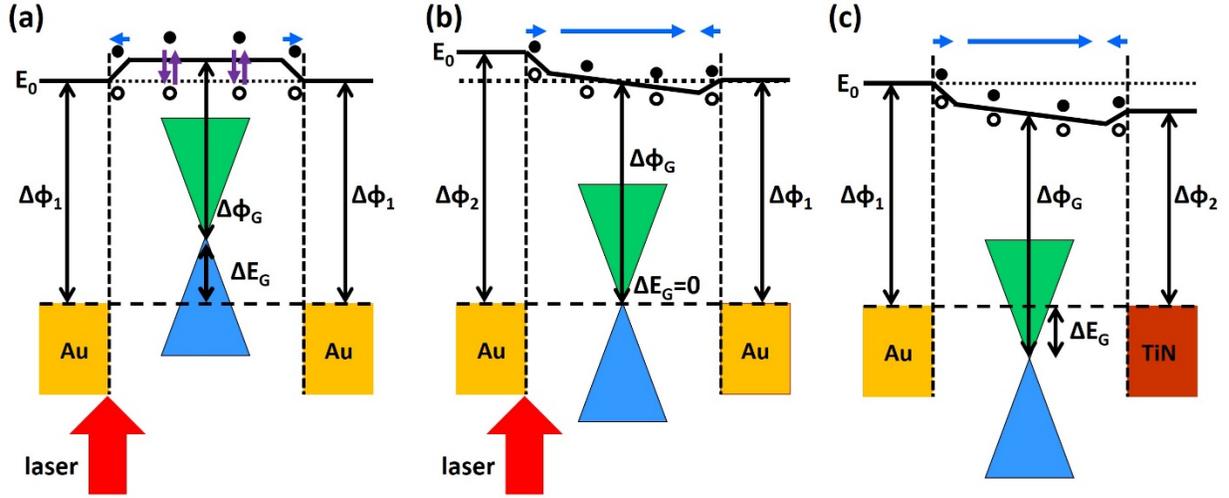

**Fig. 9.** Schematic diagrams of graphene photodetectors with different source/drain metal combinations. $\Delta\Phi_1$ is the work function of Au, and $\Delta\Phi_2$ is the work function of the asymmetric metal or Au under an applied voltage. $E_0$ is the vacuum level and $\Delta E_G$ represents the doping state of the graphene channel. The black-filled (empty) circle is an electron (hole) generated by the light absorption. Because of the internal potential generated by the work function difference of both metal contacts, the photocarriers can be more efficiently collected with the asymmetric metal contacts.

It was previously observed, that for the metal-graphene-metal arrangement (MGM), with the metal being Au, a p-type doping of graphene beneath the metal is observed that is lower than the intrinsic doping of the graphene channel. A doping induced by metal is related to a difference in the work functions of the metal and graphene ($\Phi_G$=4.5 eV) [56] which leads to charge transfer at the contact interface. Depending on the metal, a different type and doping level can be achieved [22, 55-57]. As a result, Ti ($\Phi_G$=4.3 eV) shifts the Fermi energy for $\Delta E_F$=-230 meV while Au ($\Phi_G$=4.7 eV) for $\Delta E_F$=250 eV [56]. This confirms that Ti contacts result in n-type doping of graphene while Au contacts in p-type doping [55-57]. Furthermore, a doping of graphene can be realized by injection of nonequilibrium hot electrons generated from the plasmonic structures into a nearby graphene structure [65].

The transit-time-limited bandwidth of the photodetector is given by [14, 29]:

$$f_t = \frac{3.5}{2\pi t_r} \tag{15}$$

where $t_r$ is the transit time between the metal stripe and external electrode. The proposed photodetector can operate even at the zero bias voltage as the result of a difference in Fermi level between two contacts on graphene that are doped by different metallic electrodes (Fig. 9). In consequence, the built-in electric field is created. Assuming a carrier velocity of $1.1 \cdot 10^5$ cm/s [29], and distance between electrodes of 350 nm, a single transit-time limited bandwidth exceeding 180 GHz is expected.

## 9. Conclusion

In summary, we have proposed an on-chip graphene plasmonic photodetector in an asymmetric contact arrangement able to generate a strong electronic temperature gradient across a graphene channel. As a result, an estimated electronic temperature exceeding 12000 K at power of 33 µW was calculated and responsivity exceeding 200 A/W at power below 15 µW was predicted. The light-graphene interaction was enhanced through integration of an SPP waveguide with a graphene. In consequence, over 40 % of input power was absorbed by graphene creating a confined heat source in the interface between metal and graphene. This allows a drastically enhanced PTE current across the photodetector channel. Furthermore, we have identified a regime where channel PTE can be isolated from other effects giving rise to the overall photocurrent increase.

## Author information

**Affiliations**
New York University Abu Dhabi, Saadiyat Island, PO Box 129188, Abu Dhabi, UAE
Jacek Gosciniak & Mahmoud Rasras
John Hopkins University, Baltimore, MD 21218, USA
Jacob Khurgin

**Contributions**

**Corresponding author**

Correspondence to [Jacek Gosciniak](mailto:jeckug10@yahoo.com.sg) (jeckug10@yahoo.com.sg)